\documentclass[%
 reprint,
 amsmath,amssymb,
 aps,
]{revtex4-2}

\usepackage{graphicx}
\usepackage{float}
\usepackage{dcolumn}
\usepackage{bm}
\usepackage{multirow}
\usepackage{hyperref}
\usepackage{placeins}
\usepackage{wrapfig}
\usepackage{color}

\begin{document}

\preprint{APS/123-QED}

\title{Quasiparticle Dynamics in Superconducting Quantum-Classical Hybrid Circuits}

\author{Kuang Liu$^{1,2,3}$}
\author{Xiaoliang He$^{1,2,3}$}%
\author{Zhengqi Niu$^{1,2,4}$}%
\author{Hang Xue$^{1,2}$}%
\author{Wenbing Jiang$^{1,2}$}%
\author{Liliang Ying$^{1,2}$}%
\author{Wei Peng$^{1,2,3}$}%
\author{Masaaki Maezawa$^{1,2}$}%
 \email{electronic address: masaaki.maezawa@mail.sim.ac.cn}
\author{Zhirong Lin$^{1,2,3}$}%
 \email{electronic address: zrlin@mail.sim.ac.cn}
 \author{Xiaoming Xie$^{1,2,3,4}$}%
\author{Zhen Wang$^{1,2,3,4}$}%
 \email{electronic address: zwang@mail.sim.ac.cn}
\affiliation{%
$^1$National Key Laboratory of Materials for Integrated Circuits, Shanghai Institute of Microsystem and Information Technology, Chinese Academy of Sciences, Shanghai 200050, China\\
$^2$CAS Center for Excellence in Superconducting Electronics, 865 Changning Rd., Shanghai 200050, China\\
$^3$University of Chinese Academy of Sciences, Beijing 100049, China\\
$^4$ShanghaiTech University, Shanghai 201210, China\\
}%


\date{\today}

\begin{abstract}
Single flux quantum (SFQ) circuitry is a promising candidate for a scalable and integratable cryogenic quantum control system. However, the operation of SFQ circuits introduces non-equilibrium quasiparticles (QPs), which are a significant source of qubit decoherence. In this study, we investigate QP behavior in a superconducting quantum-classical hybrid chip that comprises an SFQ circuit and a qubit circuit. By monitoring qubit relaxation time, we explore the dynamics of SFQ-circuit-induced QPs. Our findings reveal that the QP density near the qubit reaches its peak after several microseconds of SFQ circuit operation, which corresponds to the phonon-mediated propagation time of QPs in the hybrid circuits. This suggests that phonon-mediated propagation dominates the spreading of QPs in the hybrid circuits. Our results lay the foundation to suppress QP poisoning in quantum-classical hybrid systems.

\end{abstract}

\maketitle


\section{\label{sec:level1}INTRODUCTION}

Superconducting qubits are promising for constructing large-scale quantum processors thanks to the advantages of the macroscopic quantum nature of superconductivity and scalability of solid-state integrated circuits \cite{kjaergaard2020superconducting}.\;It has been reported that the integration of tens or even hundreds physical qubits was realized with high-fidelity control and readout \cite{arute2019quantum,gong2021quantum,ball2021first}. However, it is still challenging to further scale up superconducting quantum processors to more than thousands of qubits in a dilution refrigerator.\;The difficulty is associated mainly with the conventional microwave-based control scheme using a bunch of coaxial cables between room- and low-temperature stages: the management of heat load, signal latency, and hardware overhead becomes harder with increasing the number of qubits \cite{franke2019rent}.\;A potential solution is developing cryogenic quantum-classical interfaces, such as cryogenic CMOS \cite{bardin201929,pauka2021cryogenic}, superconducting single flux quantum (SFQ) circuits \cite{mcdermott2018quantum,li2019hardware,jokar2022digiq,PRXQuantum.3.010350,doi:10.1063/5.0083972,9586326}, and photonic diodes \cite{lecocq2021control}, etc. 
~\\

Among these quantum-classical interfaces, SFQ-based circuitry \cite{likharev1991rsfq,mukhanov2011energy,ying2021development} has an advantage of extremely-low power consumption and excellent compatibility to the superconducting qubits, both of which are made of superconductors, making hybridization straightforward. Theoretically, SFQ-based universal quantum gates, including both single-bit gates and two-bit gates, can be realized with a high fidelity of over 99.9 $\%$ {\cite{PhysRevApplied.2.014007,PhysRevApplied.6.024022,PhysRevApplied.19.044031}}. However, the current experimental fidelity of SFQ-based single-bit gate operations has not exceeded 98.8$\%$ \cite{PRXQuantum.3.010350,PhysRevApplied.11.014009,liu2023single}. A main cause of the fidelity reduction, which should be solved for developing practical large-scale quantum-classical systems, is attributed to quasiparticle (QP) poisoning \cite{PhysRevApplied.11.014009} commonly known as one of the main decoherence sources of superconducting qubits \cite{martinis2014ucsb,gordon2022environmental}.

In previous research, the behaviors of QPs induced by cosmic rays \cite{mcewen2022resolving}, high-power injection \cite{wang2014measurement}, propagating photon-assisted pair-breaking\cite{liu2023single}, NIS junction injection \cite{PhysRevB.96.220501}, and {SIS junction injection} \cite{iaia2022phonon} have been studied. There are two typical QP propagation mechanisms in superconductors: diffusion and propagation via medium phonons.\;Generally, when the normalized QP \linebreak density is low, most of the nonequilibrium QPs diffuse in the superconductor and then are trapped in low bandgap regions such as magnetic flux vortices or superconducting defects  \cite{wang2014measurement,PhysRevB.94.104516,taupin2016tunable}.\;When the normalized \linebreak QP density is high, QP recombination is dominant and phonon-mediated propagation is the leading propagation mechanism \cite{PhysRevB.96.220501}. A dominant QP propagation mechanism \linebreak is determined by the QP density and thus the QP generation mechanism as well as device configurations. Different QPs suppression schemes are suitable for the corresponding propagation mechanisms (e.g., normal metal traps \cite{PhysRevB.94.104516} and gap engineering {\cite{PhysRevB.100.144514,PhysRevLett.108.230509,pan2022engineering}} for diffusion-dominated QPs, phonon traps {\cite{iaia2022phonon,henriques2019phonon}} for QPs propagated by medium phonon).\enspace It is essential to clarify the QP dynamics, which has not been clear yet for the quantum-classical hybrid circuits, in order to find optimal ways of suppressing the QP poisoning for high fidelities close to the theoretical prediction. 
~\\

This paper explores the QP dynamics in a hybrid quantum-classical circuit that comprises of a DC/SFQ converter and a transmon qubit. The DC/SFQ converter serves as a QP generator, and the transmon qubit is used to monitor QP densities. We have conducted systematic \linebreak measurements of qubit relaxation rates, which are found to depend strongly on the density of QPs around the qubits.\;Our experimental results for QP densities at the qubit agree with {the timescale of phonon-mediated QP propagation model.\;The results suggest that phonon-mediated QP propagation dominates QP poisoning in quantum-classical on-chip hybrid circuits and maybe} preventing phonon propagation is more effective than capturing diffusing QPs in the superconductor for mitigating QP poisoning of the qubits in quantum-classical circuits.

\section{\label{sec:level2}EXPERIMENT AND RESULTS}

\begin{figure}[ht]
\centering
\includegraphics[width=0.5\textwidth]{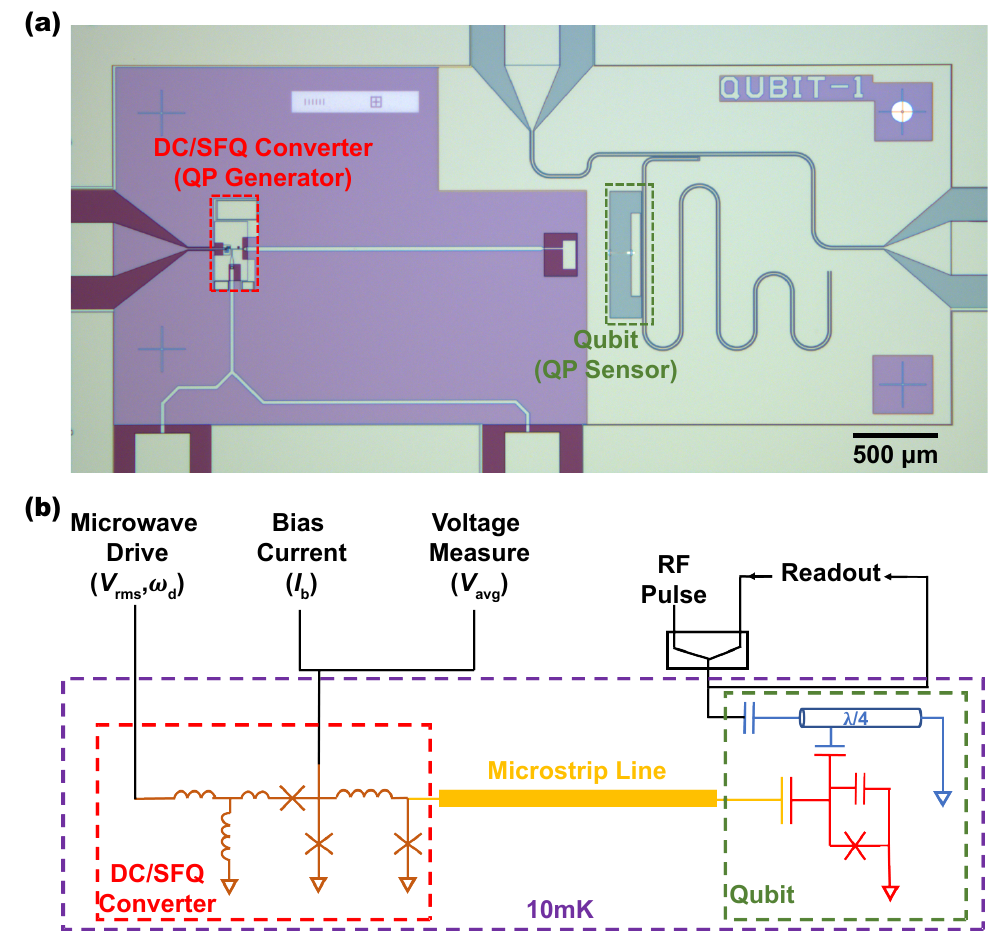}
\caption{\label{fig1}The device and experimental setup for investigating the dynamics of non-equilibrium QPs in superconducting quantum-classical circuits. \textbf{(a)} Optical micrograph of the device showing the monolithic integration of a DC/SFQ converter, microstrip line for transmitting SFQ pulse train, coupling capacitor, transmon qubit, and readout resonator. {The metal luster in the optical micrograph is the color of niobium or aluminum film, the purple part is covered with the $\mathrm{SiO_2}$ film, and the dark green is the color of the substrate.} \textbf{(b)} Schematic diagram of the circuit with the test setup. The operation of the DC/SFQ converter is monitored by measuring the DC average voltage $V_{\mathrm{avg}}$ at the output, while the qubit state is observed by measuring the scattering parameter S21 of the resonator.}
\end{figure}

The experimental setup for investigating the dynamics of QPs in superconducting quantum-classical systems is shown in Fig.\;\ref{fig1}. A classical DC/SFQ converter \cite{likharev1991rsfq} and a transmon qubit which are monolithically integrated on a chip are connected via a superconducting microstrip line and a coupling capacitor. The DC/SFQ converter generates the excess QPs along with the SFQ-voltage pulses ($\int V\left ( t \right )\, \mathrm{d}t=\Phi _{0} \equiv h/2e$) produced by switching Josephson junctions during the operation of the SFQ-based classical circuits. The amount of introduced QPs can be adjusted by changing the frequency or duration of SFQ circuit operation. The excess nonequilibrium QPs propagate, recombine, partially reach the vicinity of the qubit, and ultimately enhance the qubit relaxation. The relaxation rate of qubit can be measured using $T_1$ measurements. By adjusting the timing of the qubit relaxation rate measurement, we can monitor the evolution of the QP density and study QP dynamics in this superconducting quantum-classical hybrid circuits.

\subsection{\label{sec:level2_1}Device Fabrication and Measurement}

In Fig.$\;$\ref{fig1}(a), we show a micrograph of the superconducting quantum-classical hybrid circuit device.$\,\,$The device consists of a DC/SFQ converter on the left, which can generate a definite SFQ pulse sequence, and a qubit on the right, which is capacitively coupled to the SFQ pulse via a 2-mm-long superconducting microstrip line and a coupling capacitor.\;To reduce the static power consumption of the SFQ circuit, we eliminated the on-chip bias current shunt resistor.$\,\,$Our device fabrication process involved stacking Nb/Al-$\mathrm{AlO}_\mathrm{x}$/Nb multilayers for in-situ deposition, oxidation, and etching of SFQ circuit elements, including Josephson junctions and inductors, along with the qubit capacitor, readout resonator, and microwave ground on a high-resistance silicon substrate.$\;$We then deposited and patterned the first insulating layer ($\mathrm{SiO}_2$), resistance layer (Pd) for shunt resistance in the SFQ circuit, the second insulating layer ($\mathrm{SiO}_2$), and superconducting layer (Nb) for circuit wiring, including the microstrip line, layer by layer. The insulating layer near the qubit capacitor and the readout resonator was removed by buffered oxide etchant, followed by shadow evaporation and lift-off of Al/$\mathrm{AlO}_\mathrm{x}$/Al Josephson junctions for the qubit.

The fabricated chip was loaded into a mixing chamber of a dilution refrigerator at base temperature below 10 mK. Fig.\;\ref{fig1}(b) shows the measurement setup of the superconducting quantum-classical hybrid circuit.\;With a proper bias current $I_\mathrm{b}$, the DC/SFQ converter was driven by a microwave pulse and the operation was monitored by measuring the DC average of the output voltage $V_\mathrm{avg}$. The voltage amplitudes of the microwave drive $V_\mathrm{rms}$ applied to the DC/SFQ converter were determined at room temperature so that the decay during the transmission to the low temperature stage was not calibrated. The attenuation between the room temperature electronics and the drive port of the SFQ circuit is approximately -48 dB.\;When the bias current is fixed, the interval and duration of the SFQ pulse train can be modulated by the driving microwave pulse. The transmon qubit state was observed by measuring the scattering parameter S21 of the coupled readout resonator.

\begin{figure}[b]
\centering
\includegraphics[width=0.45\textwidth]{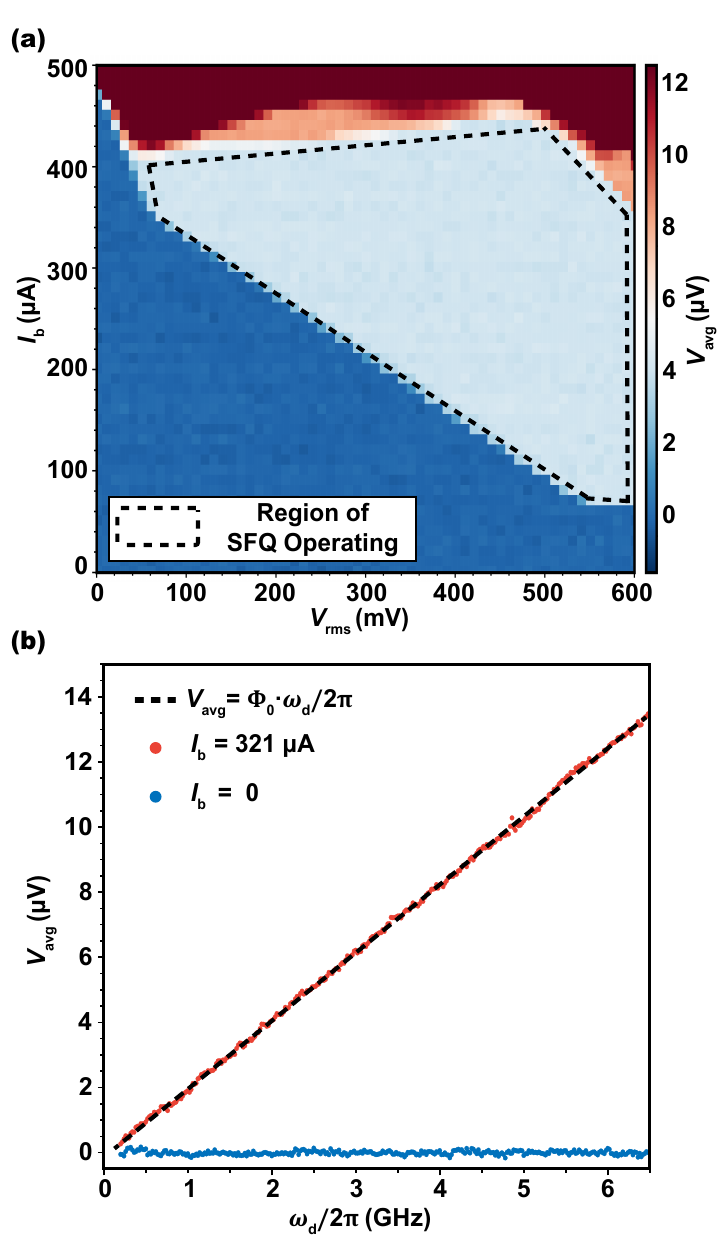}
\caption{\label{fig2}Characterization of the DC/SFQ converter in the quantum-classical circuits. \textbf{(a)} Measured average voltage $V_{\mathrm{avg}}$ as functions of the bias current $I_{\mathrm{b}}$ and drive amplitude $V_{\mathrm{rms}}$ at a fixed drive frequency ${\omega_{\mathrm{d}}}/{2\pi}=2\; \mathrm{GHz}$. The pale blue area enclosed by the dashed line corresponds to the operating region. \textbf{(b)} Measured average voltage $ V_{\mathrm{avg}}$ as a function of drive frequency $\omega_{\mathrm {d}}$ for fixed $I_\mathrm{b}$ = 321 $\mathrm{\mu A}$ and $V_{\mathrm{rms}}$ = 300 mV. The proportional relationship between $V_{\mathrm{avg}}$ and ${\omega_{\mathrm{d}}}$ confirms the correct operation.}
\end{figure}

\subsection{\label{sec:level2_2}Characterization of Superconducting Classical Circuits}
We first characterized the DC/SFQ converter in order to determine stable operating points of the DC bias current $I_\mathrm{b}$ and microwave-drive amplitude $V_\mathrm{rms}$ for the well-controlled QP generation. The operation of the DC/SFQ converter driven at $\omega_{\mathrm{d}}$ is verified by measuring the average output voltage $V_\mathrm{avg}$: if $V_\mathrm{avg} =\Phi _0\cdot \omega _\mathrm{d}/2\pi $ is obtained, the correct operation is guaranteed via AC Josephson effect\;\cite{RevModPhys.36.223}.\;Fig.\;\ref{fig2}(a) shows the measured $V_\mathrm{avg}$ after the offset-voltage calibration as functions of $I_\mathrm{b}$ and $V_\mathrm{rms}$. Here the drive frequency was fixed at ${\omega_{\mathrm{d}}}/{2\pi}=2\; \mathrm{GHz}$. The dark blue and red areas indicate error operations of deficit and excess SFQ-voltage pulse generation, respectively.\;The pale blue area enclosed by the dashed line is the correctly operating region determined with a criterion of ±3.6$\,\%$ between the measured and theoretical $V_\mathrm{avg}$. We also confirmed sufficiently wide operating regions similar to Fig.\;\ref{fig2}(a) at different drive frequencies ${\omega_{\mathrm{d}}}/{2\pi}$ ranging 0.2 GHz to 6.5 GHz.\;Fig.\;\ref{fig2}(b) shows the dependence of the measured $V_\mathrm{avg}$ on ${\omega_{\mathrm{d}}}$ for fixed $I_\mathrm{b}$ = 321 $\mathrm{\mu A}$ and $V_{\mathrm{rms}}$ = 300 mV. The slope of the $V_\mathrm{avg}$ with the change in ${\omega_{\mathrm{d}}}/{2\pi}$ is 2.065$\,\pm\, 0.003\;\mathrm{ \mu V/GHz}\approx \Phi _0$, confirming the proportional relationship and thus the correct operation over a wide frequency range.

\begin{figure*}[htb]
\centering
\includegraphics[width=0.9\textwidth]{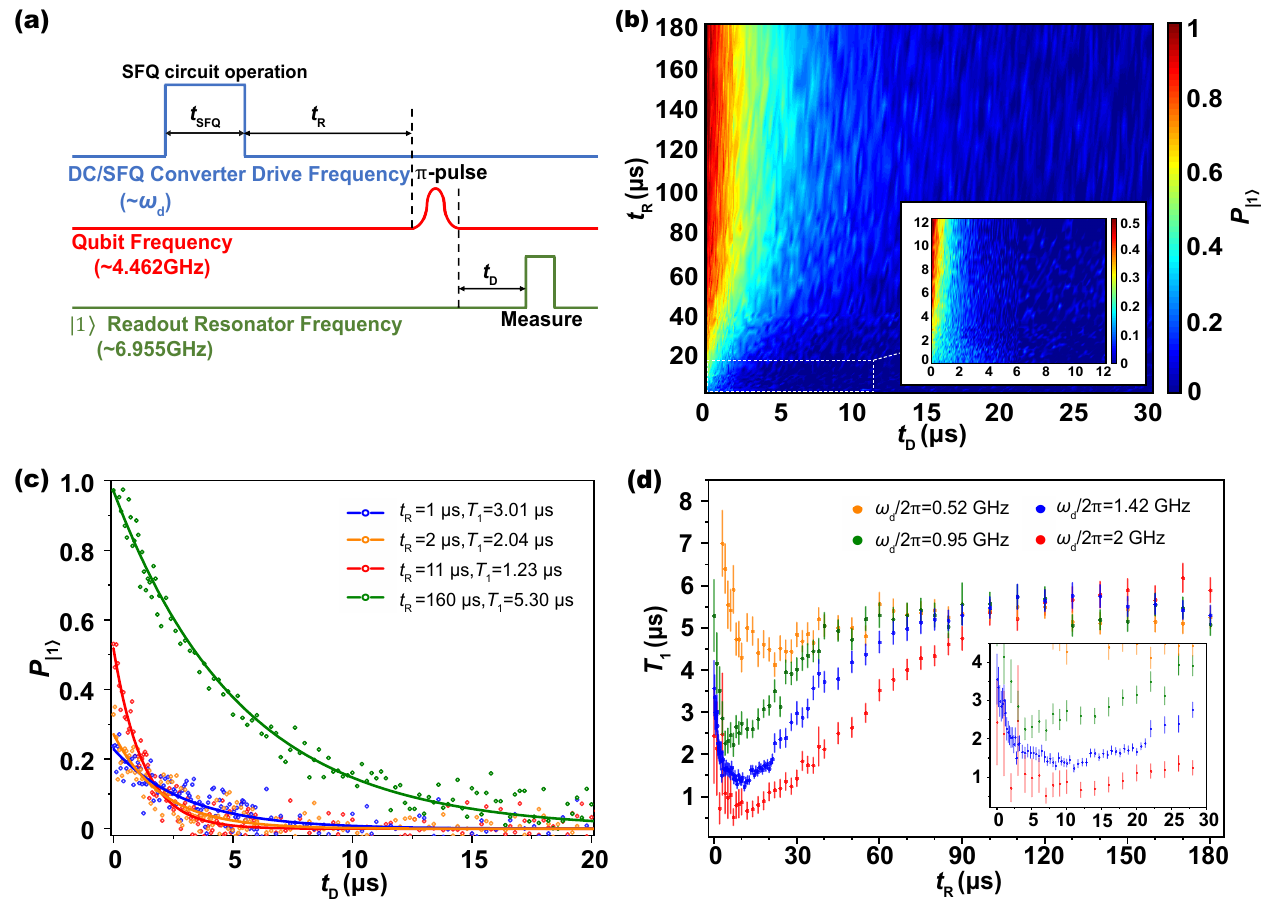}
\caption{\label{fig3} Measurement of SFQ-circuit-induced qubit relaxation time evolution. \textbf{(a)} Pulse sequences for measuring qubit relaxation rate after SFQ operation. After 25 $\mathrm{\mu s} $ SFQ operation (much longer than $\pi $ pulse and measurement pulse) with variable recovery time $t_{\mathrm{R} }$, an RF $\pi $ pulse excites qubit from $ \left | 0 \right \rangle $ to $ \left | 1  \right \rangle $, followed by a readout pulse after another variable delay $t_{\mathrm{D} }$. \textbf{(b)} Qubit $ \left | 1  \right \rangle $population as a function of delay and recovery times at $\omega_{\mathrm{d}}/2\pi=$\,1.42\,GHz. \textbf{(c)} Original qubit $ \left | 1  \right \rangle $ population data in relaxation for different
 recovery time ($t_{\mathrm R}$ = $\mathrm{1,\,2,\,11,\,160}$ $\,\mathrm{\mu s}$), at $\;\omega_{\mathrm{d}}/2\pi=$\,1.42\,GHz. Both (b) and (c) show that the qubit relaxation rate first increases and then decreases.  \textbf{(d)}  Extracted $T_{1}$ as a function of recovery time $t_{\mathrm{R} }$ for different SFQ driving frequencies $\omega _{\mathrm{d} }$ .   }
\end{figure*}

\subsection{\label{sec:level2_3}Evolution of Qubit Relaxation Time }
In the measurement setup, as shown in Fig.\;\ref{fig1}(b), an RF drive to qubit can be applied via the readout resonator port, enabling an independent qubit drive, in addition to the SFQ pulse.\;Prior to using transmon qubit as a QP sensor, it was individually characterized by applying an RF drive using room-temperature electronics, without any bias current or microwave drive applied to the DC/SFQ converter.\;The measured eigenfrequency and relaxation time $T_1$ of the transmon qubit were $\sim $4.462 GHz and  {$6.0\,(\pm\,0.2)\,\mathrm{\mu s}$}, respectively. 

~\

To investigate the changes in qubit relaxation time following SFQ circuit operation, we employed a pulse sequence depicted in Fig.\;\ref{fig3}(a) with a repetition period of 500 $\mathrm{\mu s}$.\;The DC/SFQ converter was driven at detuned frequencies for a duration of $t_{\mathrm{SFQ}} = 25\, \mathrm{\mu s}$ to generate an off-resonant SFQ pulse train. {The average number of SFQ pulses required to complete a gate operation is {strongly dependent on} the coupling capacitance of the SFQ pulses to the qubit. A typical SFQ-based X gate in Ref.\cite{PhysRevApplied.11.014009} contains 46 SFQ pulses. So, {the SFQ circuit operation duration} is equivalent to performing 280$\sim$1080 SFQ-based qubit gate operations for the driving frequency ${\omega_{\mathrm{d}}}/{2\pi}$ range of 0.52 to 2 GHz.} Nonequilibrium QPs were induced during this interval.\,The DC bias current $I_{\mathrm{b}}$ and microwave-drive amplitude $V_{\mathrm{rms}}$ were fixed at 321 $\mathrm{\mu A}$ and 300 mV, respectively, at room temperature. Following a recovery time $t_{\mathrm{R}}$, the qubit was excited using a $\pi$ pulse from room-temperature electronics, after which the qubit relaxation time was measured.

~\

Fig.\;\ref{fig3}(b) and (c) present measurement results showing that the SFQ circuit operation enhances the qubit decoherence and induces QP poisoning at ${\omega_{\mathrm{d}}}/{2\pi}=1.42\;\mathrm{GHz}$. The qubit relaxation time decreases up to a dip at $t_{\mathrm{R}}\approx\,11 \mathrm{\mu s}$. As the recovery time is prolonged, it eventually returns to the level without SFQ circuit operation. The relaxation time $T_1$ of the qubit is extracted with various recovery times after SFQ circuit operation, as shown in Fig.\;\ref{fig3}(d). $T_1$ drops sharply to $\sim1.23\,\mathrm{\mu s} $ and then recovers gradually with the recovery time. Similar trends in $T_1$ are also observed in different SFQ driving frequencies (see Fig.\;\ref{fig3}(d)) or driving durations (see Appendix \ref{app:A}). Although the quantity of QPs reaches its peak after the SFQ circuit operation, the qubit relaxation time does not immediately drop to its minimum. This implies that propagation of QPs from the SFQ circuit to the qubit takes microseconds. And the driving frequency has an impact on the minimum of the relaxation time of the qubit, with a higher frequency resulting in a shorter minimum relaxation time.\;This trend suggests that a greater number of QPs are able to reach the qubit when the Josephson junction is switched more frequently. However, it's worth noting that even after a long SFQ pulse train, the total population of both the $ \left | 0  \right \rangle $ and $ \left | 1  \right \rangle $ states is less than 1 (see Appendix \ref{app:B}). This means that the $\pi$ pulse used to prepare the qubit to the $ \left | 1  \right \rangle $ state is not always effective. While we plan to investigate the reasons for this behavior in future work, we can still use the relaxation of the $ \left | 1  \right \rangle $ state population to determine the $T_1$ of the qubit.

\section{\label{sec:level3}DISCUSSION AND ANALYSIS}

To describe the evolution of QP density, we calculated $x_{\mathrm{QP,qubit} }$, the average normalized {excess} QP density near the qubit with the measured $T_1$ data, using the equation \cite{wang2014measurement,PhysRevB.84.064517} 
\begin{equation}
x_{\mathrm{QP,qubit} } =\frac{\Gamma \left (  t \right )-\Gamma _{0}  }{C}, \label{equ1}
\end{equation}
 where $C\,$=$\sqrt{2\omega _{\mathrm{q} }\Delta /\pi ^{2}\hbar} $, $\omega _{\mathrm{q} }$($\approx 2\pi \cdot 4.462\,\mathrm{GHz} $)  is  the  
\noindent angular eigenfrequency of the transom qubit, $\Delta ( \approx 0.18\,\mathrm{meV} )\; $ is the gap energy of the qubit Josephson junction electrodes, $\Gamma \left ( t  \right )=1/T_{1} $ is the measured qubit relaxation rate, $\Gamma _{0} $ represents the qubit relaxation rate without nonequilibrium QP introduced. {Note that because $T_1$ measurement itself takes time, the $x_{\mathrm{QP,qubit} }$ extracted using Eq. (\ref{equ1}) is the equivalent average excess QP density during the $T_1$ measurement, which can be regarded as the average density of QP over a period of time ($\sim T_1$) after $t_\mathrm{R}$. The qubit $T_1$ varies in the microsecond range, therefore $x_{\mathrm{QP,qubit}}$ extracted is sufficient to describe the SFQ-induced QP dynamics on timescale above the microseconds level (see Appendix \ref{app:C} for details).} Fig.\;\ref{fig4} illustrates the evolution of $x_{\mathrm{QP,qubit}}$ calculated by using the experimentally determined $T_1$ at the drive frequency $\omega _{\mathrm{d} } /2\pi =1.42\,\mathrm{GHz}$. We observe a rapid increase of $x_{\mathrm{QP,qubit}}$ after the SFQ circuit generates excess QPs, followed by a slow decay. 

\begin{figure}[t]
\centering
\includegraphics[width=0.45\textwidth]{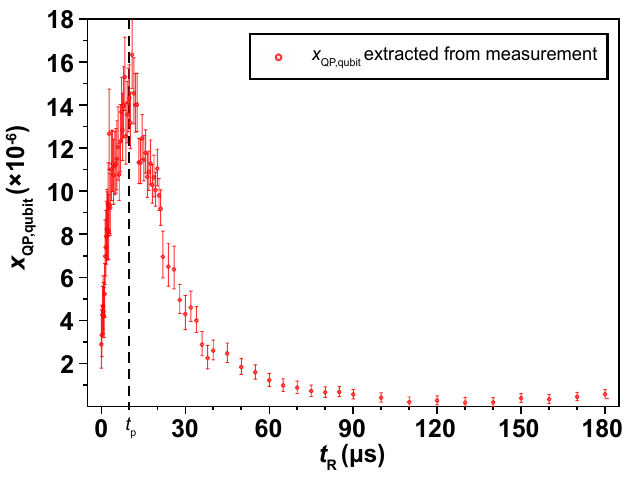}
\caption{\label{fig4} Evolution of the normalized QP density $x_{\mathrm{QP,qubit}}$ near the qubit as a function of recovery time $t_\mathrm{R}$ after SFQ circuit operation at $\omega _{\mathrm{d}}/2\pi = 1.42\;\mathrm{GHz} $. The red circles with error bars denote the values extracted from the measured qubit relaxation rate. { $x_{\mathrm{QP,qubit}}$ reaches its peak at $t_{\mathrm{R}}\approx t_{\mathrm{p}}$, which is indicated by the vertical dashed line.}}
\end{figure}
\begin{figure}[b]
\centering
\vspace{0.1 cm}
\includegraphics[width=0.47\textwidth]{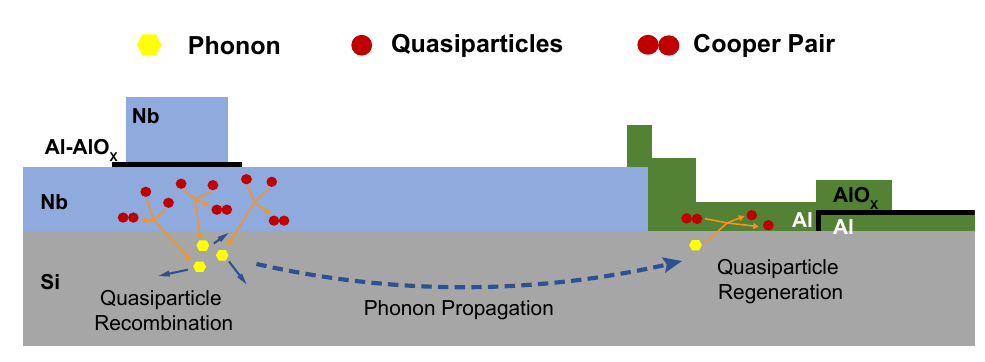}
\caption{\label{fig5} Schematic of the phonon-mediated propagation of QPs in a superconducting quantum-classical hybrid circuit.}
\end{figure}

{When the operation of the SFQ circuit ends ($t_{\mathrm{R}} = 0$), the QP density near the SFQ circuit accumulates to its peak, but the extracted $x_{\mathrm{QP,qubit}}$ reaches its peak at $t_{\mathrm{R}}=t_{\mathrm{p}} \approx 10\, \mathrm{\mu s}$. The timescale of such lag $t_{\mathrm{p}}$ is strongly dependent on the mechanism by which the SFQ-induced QPs propagate to the vicinity of the qubit. We discuss the different QP propagation mechanisms that may dominate $x_{\mathrm{QP,qubit}}$ evolution below:} 

 {(i) Microwave photons with frequencies above $2\Delta/h$ can be absorbed by Cooper pairs and produce QPs {\cite{PhysRevLett.123.107704}}. So, photon-assisted QPs may be excited near the qubit during SFQ circuit operation {due to the ultrahigh spectral components of SFQ pulses}. Considering the velocity of photons is the speed of electromagnetic waves ($\sim10^8$ m/s), the photon-assisted QPs reach the vicinity of the qubit ($\sim\,$2.5 mm) in only a few tens of picoseconds, which is much faster than the observed $t_{\mathrm{p}}$,} suggesting that photon-assisted QPs do not play a dominant role in $x_{\mathrm{QP,qubit}}$ evolution. 

  {(ii) The QPs introduced by the SFQ circuit operation can directly diffuse to the vicinity of the qubit through the superconducting thin film. According to the diffusivity of QP in niobium ($D_\mathrm{QP}\,=\,1.2\;{{\mathrm{cm}}^2}/{\mathrm{s}}$ \cite{doi:10.1063/1.363350}), the timescale for the same distance to diffuse is at least on the order of milliseconds, which is much slower than the observed $t_{\mathrm{p}}$. {Also}, because the diffusion of QPs in superconducting thin films is accompanied by QPs recombination and trapping,} the peak of $x_{\mathrm{QP,qubit}}$ ($\sim \mathrm{10^{-5}}$) is much larger than the QP density contributed by direct diffusion ($\ll\,\mathrm{10^{-5}}$; see Appendix \ref{app:D}).  
 
   {(iii)} The phonon-mediated QP propagation from the SFQ circuit to the vicinity of qubit is shown in Fig.\;\ref{fig5}. Non-equilibrium quasiparticles (QPs) are generated near the SFQ circuit during its operation, due to the switching of Josephson junctions (usually Nb/Al-$\mathrm{AlO_{x}}$/Nb trilayer stacked structure) and the voltage across their shunt resistors (not shown in the figure, used to eliminate hysteresis). As the local QP density increases, the SFQ-circuit-induced QPs have a greater probability of recombining into Cooper pairs and releasing phonons. These phonons propagate and scatter in the substrate, and the Cooper pairs in other superconducting regions may absorb the phonons with energy greater than $2\Delta $ and be broken into QPs. To estimate the timescale for such propagation, we treat the propagation of phonons from the SFQ circuit to the qubit vicinity as boundary-limit diffusion in the substrate \cite{iaia2022phonon}. Considering the random scattering of phonons at the top and bottom of the substrate during propagation, the process can be described by the effective diffusivity $D_\mathrm{phonon}=v_\mathrm{s}d$, where $v_\mathrm{s}$ is the speed of sound in Si substrate ($\sim 10^3 \mathrm{m/s}$) and $d = 0.625\,\mathrm{mm}$ is the substrate thickness. {The phonon diffusivity} leads to a timescale of several microseconds for phonons to diffuse from the SFQ circuit to the qubit vicinity ($\sim 2.5 \mathrm{mm}$), which is close to the $t_{\mathrm{p}}$. We also try to employ the phonon-mediated QP propagation in a simplified model (see Appendix \ref{app:E}), which {qualitatively agrees with the measured $x_{\mathrm{QP,qubit}}$ evolution}.



\section{\label{sec:level5}SUMMARY AND OUTLOOK}
In summary, we focus on the fabrication and characterization of quantum-classical hybrid circuits, with a particular emphasis on understanding the dynamics of SFQ-circuit-induced QPs through a transmon qubit sensor. We observed that {the qubit relaxation time initially decreased after the SFQ circuit operation but later recovered to its baseline value.}\;{To quantify the QP dynamics, we extrated SFQ-circuit-induced QP density in the vicinity of the qubit. The timescale of QP density evolution contradicted that of photon-assisted or direct-diffusion QP propagation but was consistent with phonon-mediated QP propagation.} These results support the hypothesis that phonon-mediated QP propagation is the dominant mechanism contributing to QP poisoning in quantum-classical on-chip hybrid circuits. 

The QP trapping rate in this device is significantly high {($\sim1/(10\,\mu s) $)}, exceeding that of the 3D transmon by two orders of magnitude \cite{wang2014measurement}.\;It is unlikely that fabricating normal metal traps on superconducting thin films or directly enhancing QP trapping would yield substantial improvements in suppressing QP poisoning.\;Therefore, in order to achieve low-temperature integrated high-fidelity qubit control via SFQ circuits, it is crucial to block or cut off medium phonon propagation in superconducting quantum-classical hybrid circuits.

\begin{acknowledgments}
We acknowledge the support from Superconducting Electronics Facility (SELF) in Shanghai Institute of Microsystem and Information Technology (SIMIT), CAS, for device fabrication.\;We thank Dr.\;Jie\;Ren, from SIMIT for providing SFQ design infrastructure.\;This work is supported in part by the National Natural Science Foundation of China (No.92065116), the Key-Area Research and Development Program of Guangdong Province, China (No.2020B0303030002), the Shanghai Technology Innovation Action Plan Integrated Circuit Technology Support Program (No.22DZ1100200) and Strategic Priority Research Program of the Chinese Academy of Sciences (Grant No.XDA18000000).

\end{acknowledgments}

\appendix

\section{Qubit Relaxation Time after SFQ Operation}\label{app:A}

\begin{figure}[!htbp]
\centering
\includegraphics[width=0.38\textwidth]{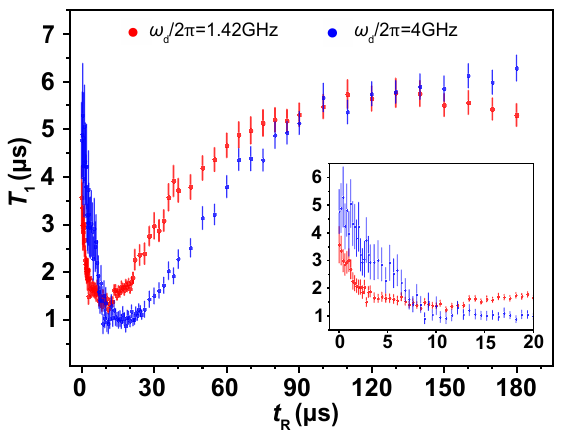}
\renewcommand{\thefigure}{6}
\caption{Extracted $T_1$ as a function of recovery time $t_{\mathrm{R}}$ for $\omega_{\mathrm{d}}/{2\pi}=1.42\,\mathrm {GHz}$, $t_{\mathrm{SFQ}}=25\,\mathrm{ \mu s}$ and $\omega_{\mathrm{d}}/{2\pi}=4\,\mathrm {GHz}$, $t_{\mathrm{SFQ}}=8.87\,\mathrm{ \mu s}$.
\textbf{(insert)} The zoomed in view in the range of 0 to 20 $\mathrm{\mu s}$.
}\label{fig-6}
\end{figure}

\begin{figure}[!htbp]
\centering
\includegraphics[width=0.38\textwidth]{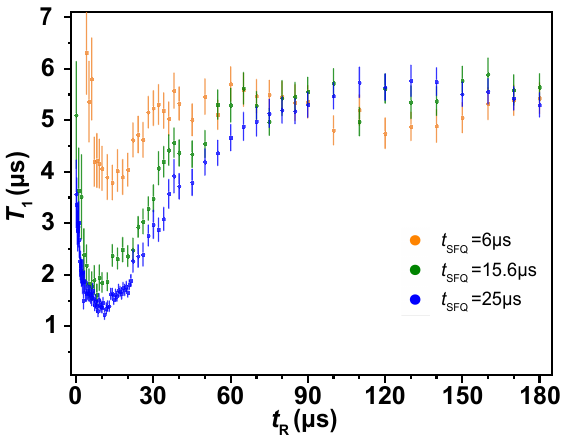}
\renewcommand{\thefigure}{7}
\caption{\label{fig7}Extracted $T_1$ as a function of recovery time $t_{\mathrm{R}}$ for variable $t_{\mathrm{SFQ}}$ when $\omega_{\mathrm{d}}$ = $2\pi\cdot 1.42\, {\mathrm{GHz}}$.
}
\vspace{-10mm}
\end{figure}

In the main text, we show that $T_1$ for the qubit initially decreases and then gradually recovers with recovery time $t_{\mathrm{R}}$ at various driving frequencies $\omega_{\mathrm{d}}$. Similar trends in $T_1$ evolution are also observed in other SFQ driving frequencies or driving durations. At a fixed bias current $I_\mathrm{b}$= 321 $\mathrm{\mu A}$ and drive amplitude $V_{\mathrm{rms}}$=\,300 mV at room temperature, the pulse sequence in Fig.\;\ref{fig3}(a) was applied to the superconducting quantum-classical hybrid circuit. Fig.\;\ref{fig-6} shows $T_1$ evolution of different driving frequencies $\omega_{\mathrm{d}}$ with the same number of Josephson junction switching ($\omega_{\mathrm{d}}\cdot t_{\mathrm{SFQ}}$) in the SFQ circuit. A higher clock frequency results in a lower dip in the qubit relaxation time. In Fig.\;\ref{fig7}, we compare the evolution of qubit relaxation time with $t_{\mathrm{R}}$ at different $t_{\mathrm{SFQ}}$. The trends are similar and qubit $T_1$ first decreases and then gradually recovers at all driving durations. And the dip in qubit relaxation time becomes lower as the SFQ circuit operates for a longer time. This can be qualitatively explained as follows: the peak density of nonequilibrium QPs accumulated near the SFQ circuit increases when there are more times of the Josephson junction being switched (higher clock frequency or longer operation time). As a result, more QPs propagate to the vicinity of the qubit, enhancing the qubit relaxation.

\begin{figure*}[tbp]
\centering
\includegraphics[width=0.77\textwidth]{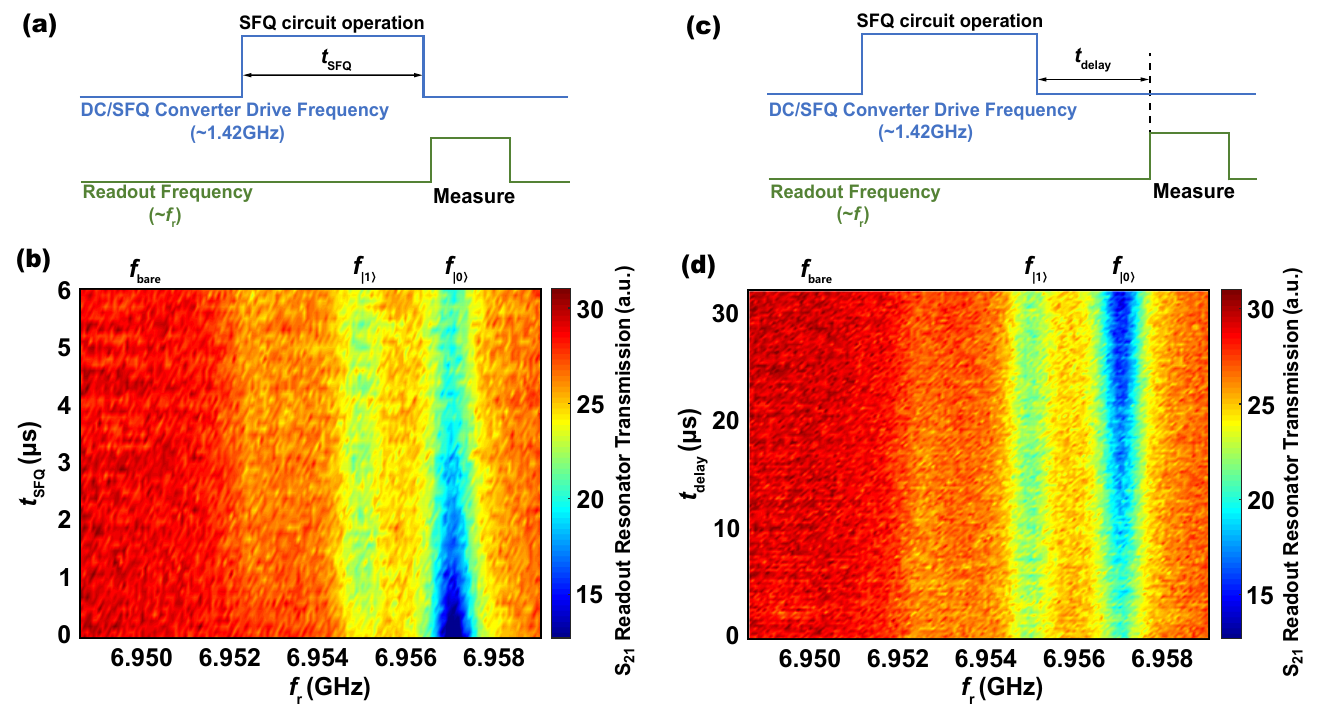}
\renewcommand{\thefigure}{8}
\caption{\textbf{(a)} Pulse sequence applied to measure qubits state after an SFQ pulse train. The transmission of the readout resonator is measured at $f_{\mathrm{r}}$. \textbf{(b)} Qubit readout resonator spectroscopy vs. DC/SFQ converter drive pulse duration $t_{\mathrm{SFQ}}$. {\textbf{(c)} Pulse sequence applied to measure qubits state with variable delay $t_\mathrm{dalay}$ after a 25$\mu\mathrm{s}$ SFQ pulse train. The transmission of the readout resonator is measured at $f_{\mathrm{r}}$. \textbf{(d)} Qubit readout resonator spectroscopy vs. delay between SFQ circuit operation and readout pulse $t_\mathrm{dalay}$.} The frequencies corresponding to the readout resonator when the state of qubit is $\left | 0 \right \rangle $, $\left | 1 \right \rangle $, and punch out have been marked.
}\label{fig8} 
\end{figure*}

\section{Qubit State Population after A Long SFQ Pulse Train}\label{app:B}

As shown in Fig. \ref{fig8} (a) and (b), we measured the qubit state when it was driven by an SFQ pulse train at an off-resonant frequency for a long time. We find that as the driving time $t_{\mathrm{SFQ}}$ increases, the qubit population at $\left | 0 \right \rangle $ tends to decrease because the qubit is excited to other states. {As shown in Fig. \ref{fig8} (c) and (d), the qubit gradually relaxes to $\left | 0 \right \rangle $ after the SFQ operation.} Therefore, when the $\pi$ pulse is applied to the qubit immediately after the SFQ operation, the qubit cannot be completely excited to the $\left | 1 \right \rangle $ state. {As the recovery time prolongs, the population of $\left | 1 \right \rangle $ after the $\pi$ pulse also gradually increases to 1.}

\section{Extraction of Quasiparticle Density Near Qubit}\label{app:C}

{The $x_{\mathrm{QP,qubit}}$ extracted using $x_{\mathrm{QP,qubit}}=(\Gamma  ( t )-\Gamma _{0})/{C}$ is a kind of average QP density during the $T_1$ measurement, which can be regarded as the average density of QP over a period of time $t_\mathrm{avg}$ ($\sim\,T_1$) after $t_\mathrm{R}$. To intuitively illustrate the validity of the $x_{\mathrm{QP,qubit}}$ evolution extracted in this way to analyze the QP propagation mechanism, we compare (i) the $x_{\mathrm{QP,qubit}}$ extracted with a very short measurement time ($t_\mathrm{avg}\,=\,0$) in an ideal measurement, and (ii) the $x_{\mathrm{QP,qubit}}$ extracted with several microsecond measurement times ($t_\mathrm{avg}=3,\ 6,\ 12,\ 18\ \mu s$) in a real experiment. We assume that the gray curve ($t_\mathrm{avg}\,=0$) in the Fig. \ref{fig9} below, which is similar to the trend of the extracted $x_{\mathrm{QP,qubit}}$ evolution in the maintext, is the real QP density evolution. The trends and timescales ($\sim$several microseconds) of $x_{\mathrm{QP,qubit}}$ evolution vary little over all values of $t_\mathrm{avg}$ listed.}
\begin{figure}[htbp]
\centering
\includegraphics[width=0.4\textwidth]{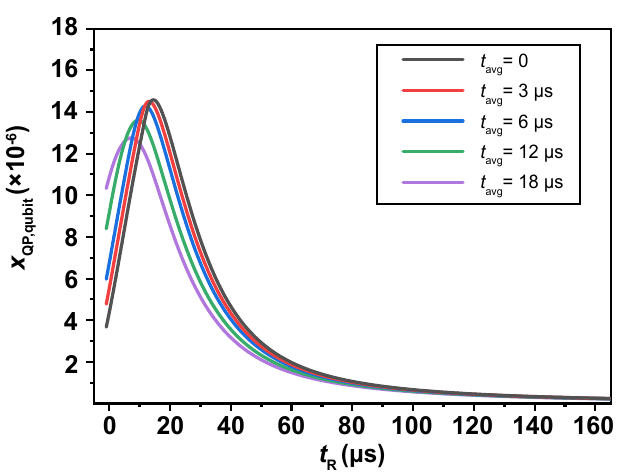}
\renewcommand{\thefigure}{9}
\caption{Effect of measurement time on the extracted $x_{\mathrm{QP,qubit}}$ evolution
}\label{fig9}
\end{figure}

\section{Diffusion of Quasiparticles in Superconducting Quantum-Classical Hybrid Circuits}\label{app:D}

\begin{figure}[tb]
\centering
\includegraphics[width=0.4\textwidth]{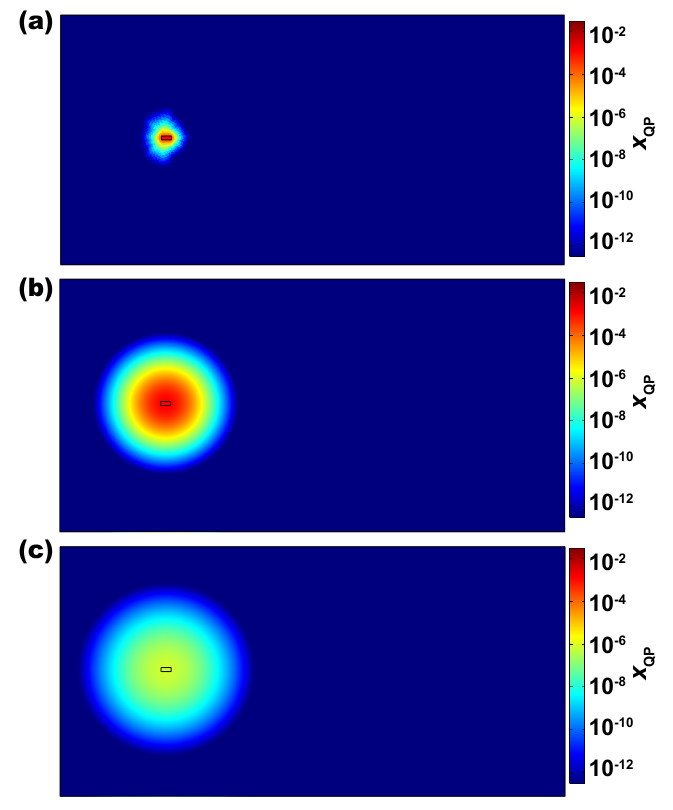}
\renewcommand{\thefigure}{10}
\caption{\label{fig10}QP density distribution in superconducting quantum-classical hybrid circuit device when diffusion dominates QP propagation. In agreement with the main text, the moment of the end of QP injection is defined as $t_{\mathrm{R}}=0$, so the moment of the start of injection is $t_{\mathrm{R}}=-25\mathrm{\; \mu s}$, corresponding to \textbf{(a)}  $t_{\mathrm{R}}=-24.99\mathrm{\, \mu s}$,  \textbf{(b)}  $t_{\mathrm{R}}=52\mathrm{\, \mu s}$, \textbf{(c)} $t_{\mathrm{R}}=120\mathrm{\, \mu s}$.}
\end{figure}

Most QPs propagate diffusively when the local QP density is relatively low. The local QP density is $x_{\mathrm{QP}}=n_{\mathrm{QP}}/n_{\mathrm{CP}}$, where $n_{\mathrm{QP}}$ is the QP density and $n_{\mathrm{CP}}$ the Cooper pair density. As the local QP density increases, QPs have a greater probability of recombination, and phonon-mediated propagation becomes the leading mechanism. We calculated the diffusion equation by finite element simulation, thereby estimating the contribution of QP diffusion to the QP propagation in the device described in the paper. We set a boundary of $5\; \mathrm{mm} \times 2.5\; \mathrm{mm}$ (corresponding to the actual device size) and applied the QP dynamics equation including the diffusion term\cite{PhysRevB.94.104516,PhysRevApplied.8.064028}:
\begin{equation}\label{equD1}
\begin{aligned}
\dot{x}_{\text{QP}}(\vec{x},t_\mathrm{R})= & D_{\text{QP}}\nabla^2x_{\text{QP}}(\vec{x},t_\mathrm{R})-rx_{\text{QP}}(\vec{x},t_\mathrm{R})^2 \\
 &{-sx_{\text{QP}}(\vec{x},t_\mathrm{R})+g(\vec{x},t_\mathrm{R})} ,
\end{aligned}
\end{equation}
where the QP density $x_{\mathrm{QP}}\left(\vec{x},t_\mathrm{R}\right)$ is a function of both time and spatial coordinates; $D_{\mathrm{QP}}$ is the QP diffusion constant, which we set as the diffusion constant for QPs in niobium thin films ($D_\mathrm{QP}\,=\,1.2\;{{\mathrm{cm}}^2}/{\mathrm{s}}$) \cite{doi:10.1063/1.363350}; $r$ is the QP recombination rate, which we set as {$r=0$ to highlight the role of diffusion by neglecting QPs recombination}; $s$ is the QP trapping rate, which we set to {the estimated value of QP density decay rate in the maintext} to a uniform global value in this calculation {[$s\,=\,{1}/{(10\,\mathrm {\mu s})}$]}; and $g$ describes the generation of QP. In this section, we focus on the QP diffusion and artificially ignore the phonon-mediated propagation of QPs and the generation of QPs induced by other sources such as stray radiation. First, we injected nonequilibrium QPs into the $100\; \mathrm{\mu m} \times 40\; \mathrm{\mu m}$ area, which corresponds to the DC/SFQ converter and its surrounding area, within SFQ circuit operation duration $t_{\mathrm{SFQ}}=\,25\;\mathrm{\mu s}$. The QP generation term is $g=4\times{10}^4$ when $-t_{\mathrm{SFQ}}<t_{\mathrm{R}}\le0$, which equivalent to introducing an equal number of Cooper pairs of QPs during SFQ operations ($g\cdot t_{\mathrm{SFQ}}=1$). The amount of injected QPs is exaggerated to highlight the role of diffusion. 

\begin{figure}[t]
\centering
\includegraphics[width=0.4\textwidth]{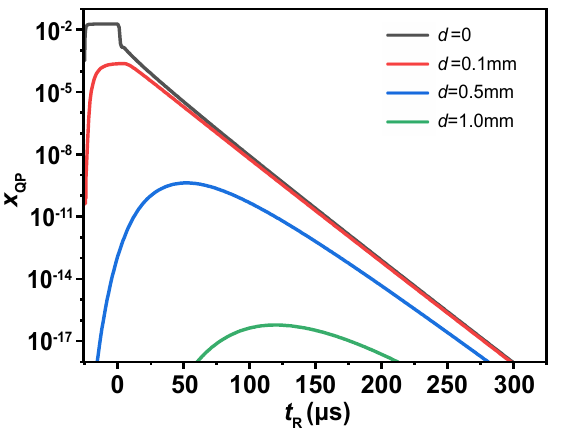}
\renewcommand{\thefigure}{11}
\caption{\label{fig11}QP density evolution for $s\,=\,{1}/{(10\,\mathrm {\mu s})}$ at different distances, $d$, from the SFQ circuit on the straight line where the SFQ circuit and qubit are located.}
\end{figure}
\begin{figure}[tbp]
\centering
\includegraphics[width=0.4\textwidth]{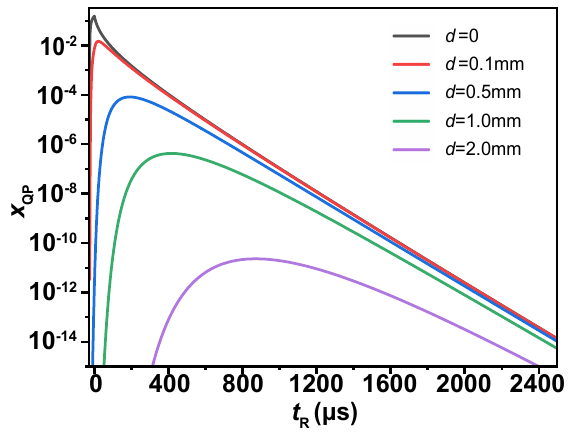}
\renewcommand{\thefigure}{12}
\caption{\label{fig12}QP density evolution for $s\,=\,{1}/{(100\,\mathrm {\mu s})}$ at different distances, $d$, from the SFQ circuit on the straight line where the SFQ circuit and qubit are located.}
\end{figure}

Then, the QPs density distribution of the device at a series of time instants is shown in Fig.\,\ref{fig10}, where (a) corresponds to the QP density distribution when the QPs have not yet diffused, and (b) and (c) correspond to the QP density distributions when the QP density peak diffuses 0.5 mm and 1 mm, respectively. Fig.\,\ref{fig11} shows the QP density evolutions at 0, 0.1, 0.5, and 1 mm away from the SFQ circuit. The moment of QP density peak arrival is rapidly delayed with increasing diffusion distance, while the peak of QP density also decreases rapidly. Even at 0.5 mm from the SFQ circuit (much smaller than the distance of $\sim$2.5 mm between the SFQ circuit and qubit in the superconducting quantum-classical hybrid circuit), the diffused QPs are no longer sufficient to have a significant enhancement on the qubit relaxation in this device. To further exaggerate the contribution of diffusion to QP propagation, we performed similar simulations with a tenfold reduction in trapping rate {[$s\,=\,{1}/{(100\,\mathrm {\mu s})}$]}. The results shown in Fig.\,\ref{fig12} show that the QPs by direct diffusion are still insufficient to be observed in this experiment. This indicates that the QP diffusion is local, and it is not the dominant propagation mechanism in this device.

\section{Model of Phonon-mediated Propagation}\label{app:E}

{As discussed in the main text, the measured peak lag of QP density near the qubit $x_{\mathrm{QP,qubit}}$ is consistent on the timescale with phonon-mediated QP propagation. To illustrate the process more concretely, we made a semi-quantitative model based on phonon-mediated propagation.} The local QP dynamics equation can be described as \cite{wang2014measurement,PhysRevB.96.220501,PhysRevB.94.104516},
\begin{equation}
\dot{x}_{\mathrm{QP}, i}=-r x_{\mathrm{QP}, i}^2-s_i x_{\mathrm{QP}, i}+g_i \label{equ2},
\end{equation}
where the index $i\in \left \{ \mathrm{SFQ,qubit}  \right \} $ denotes the location where the parameters related to the position are averaged; $x_{\mathrm{QP} ,i} = n_{\mathrm{QP} ,i}/n_{\mathrm{CP} ,i}$,  $n_{\mathrm{QP} ,i}$ is QPs density, $n_{\mathrm{CP} ,i}$ is Cooper pairs density; $r$ is the QP recombination rate; $s_{i}$ is the effective trapping rate; $g_{i}$ represents the rate of QP generation such as QP injection or pair-breaking. Fig.\;\ref{fig5} shows the process of the SFQ-induced non-equilibrium QPs reaching vicinity of qubit by phonon-mediated propagation. Since the QP directly\linebreak diffused to the qubit vicinity is far less than the total amount of QP propagated there. The SFQ-circuit-induced QPs have a greater probability of recombining into Cooper pairs and releasing phonons instead of being trapped directly ($r\cdot x_{\mathrm{QP,SFQ} }^{2} \gg s_{\mathrm{SFQ} } \cdot x_{\mathrm{QP,SFQ} }$). These phonons propagate and scatter in the substrate, and the Cooper pairs in other superconducting regions may absorb the phonons with energy greater than $2\Delta $ and be broken into QPs, which is the dominant source of non-equilibrium QP generation near the qubit.

 Thus, we first calculate the QP density near the SFQ circuit. The QP generation rate $g$ can be described separately in the two cases of SFQ circuit operation and non-operation by the following equation:
\begin{equation}
g(t_{\mathrm{R}}) =
\begin{cases}
g_{_\mathrm{SFQ}},\;\;\;\;\;\;\;\;\;\;\;\;  
\text{if $\,-t_{\mathrm{SFQ}}\leq t_{\mathrm{R}}<0 $,} \vspace{1ex}\\
0. \;\;\;\;\;\;\;\;\;\;\;\;\;\;\; \;\; 
\text{if $\;{t_{\mathrm{R}}}\ge  0$ ,}\vspace{1ex}
\end{cases}
\label{equ3}
\end{equation}
During the SFQ circuit operation ($\,-t_{\mathrm{SFQ}}\leq t_{\mathrm{R}}<0 $), the QP generation rate is a constant related to the driving frequency of the SFQ circuit. Outside the operating time of the SFQ circuit, we approximately consider the generation rate to be 0, because other QP sources (e.g., cosmic rays) are not sufficient to induce significant qubit relaxation enhancement in the experiment. We solve Eq. \ref{equ2} under the condition that the amount of QPs introduced by the SFQ operation is large enough for the QP recombination to dominate the QP relaxation ($r\cdot x_{\mathrm{QP,SFQ} } ^2\gg s_{_\mathrm{SFQ} }\cdot  x_{\mathrm{QP,SFQ} }$). The QP density near the SFQ circuit can be expressed as
\begin{equation}
{x}_{\mathrm{QP, SFQ}} (t_{\mathrm{R}}) =
\begin{cases}
\sqrt{\frac{g_{_\mathrm{S F Q}}}{r}}\left(\tanh\left(t_{\mathrm{R}}+t_{\mathrm{SFQ}}\right)\sqrt{g_{_\mathrm{S F Q}}r}\right),\\
\;\;\;\;\;\;\;\;\;\;\;\;\;\;\;\;\;\;\;\;\;\;\;\;\;\;\;\;\;\;\;\;\;\;  \text{if $\,-t_{\mathrm{SFQ}}\leq t_{\mathrm{R}}<0 $,} \vspace{1ex}\\
1/\left(r t_{\mathrm{R}}+\left(\sqrt{\frac{g_{_\mathrm{S F Q}}}{r}}\tanh\sqrt{g_{_\mathrm{S F Q}}r} t_{\mathrm{SFQ}}\right)^{-1}\right),\\
\;\;\;\;\;\;\;\;\;\;\;\;\;\;\; \;\;\;\;\;\;\;\;\;\;\;\;\;\;\;\;\;\;\;\;\;\;\;\;\;\;\;\;\;\;\;\;\; \text{if $\;{t_{\mathrm{R}}}\ge  0$ .}\vspace{1ex}
\end{cases}
\label{equ4}
\end{equation}
Then we can obtain the QP recombination rate accompanied by phonon emission in the vicinity of the SFQ circuit as $ r\cdot\left(x_{\mathrm{QP,SFQ}}(t_{\mathrm{R}})\right)^2 $. 

\begin{figure}[t]
\centering
\renewcommand{\thefigure}{13}
\includegraphics[width=0.4\textwidth]{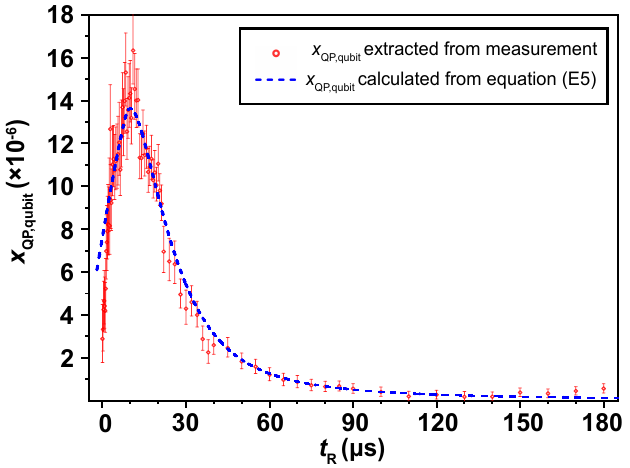}
\caption{ Evolution of the normalized QP density $x_{\mathrm{QP,qubit}}$ near the qubit as a function of recovery time $t_\mathrm{R}$ after SFQ circuit operation at $\omega _{\mathrm{d}}/2\pi = 1.42\;\mathrm{GHz} $. The red circles with error bars denote the values extracted from the measured qubit relaxation rate. The dashed line represents the QP density calculation results based on the phonon-mediated QP propagation model.}\label{fig13}
\end{figure}

These phonons propagate and scatter in the substrate. Then, the Cooper pairs in other superconducting regions may absorb the phonons with energy greater than the superconducting bandgap and be broken into QPs. Since only a fraction of the phonons can reach the qubit area and be absorbed by the Cooper pair{,} we denote the probability of such a phonon completing the above process as $\alpha$, and the time to complete a phonon-mediated propagation as $t_{\mathrm{p}}$. Then the rate of QP generation induced by phonon-mediated propagation in the qubit area ($g_{\mathrm{qubit}}$) can be expressed as $\alpha r\cdot\left(x_{\mathrm{QP,SFQ}}(t_{\mathrm{R}}-t_{\mathrm{p}})\right)^2$,that is
\begin{equation}
g_{\mathrm{qubit} } =
\begin{cases}
\alpha {g}_{\mathrm{_{SFQ}}}\left(\tanh \left(t_{\mathrm{R}}-t_{\mathrm{p}}+t_{\mathrm{SFQ}}\right) \sqrt{{ g}_{\mathrm{_{SFQ}}} r}\right)^2\vspace{1ex},\\  
\;\;\;\;\;\;\;\;\;\;\;\;\;\;\; \;\;\;\;\;\;\;\;\;\; \;\;\;\;\;\;\; \;\;\;\;\;\;\;\;\;
\text{if $\,-t_{\mathrm{SFQ}}+t_{\mathrm{p}}\le{t_{\mathrm{R}}}< t_{\mathrm{p}} $,} \vspace{1ex}\\

\alpha r\left[r\left(t_{\mathrm{R}}-t_{\mathrm{p}}\right)+\left(\sqrt{\frac{g_{\mathrm{_{SFQ}}}}{r}} \tanh \sqrt{g_{\mathrm{_{SFQ}}} r} t_{\mathrm{SFQ}}\right)^{-1}\right]^{-2}\vspace{1ex},\\ 
\;\;\;\;\;\;\;\;\;\;\;\;\;\;\; \;\;\;\;\;\;\;\;\;\; \;\; \;\;\;\;\;\;\;\;\;\; \;\;\;\;\;\;\; \;\;\;\;\;\;\;\;\; \;\;\;\;\;\;\;\;\;\;
\text{if $\,{t_{\mathrm{R}}}\ge  t_{\mathrm{p}}$ .}\vspace{1ex}
\end{cases}
\label{equ5}
\end{equation}
The non-equilibrium QP density near the qubit is relatively low, the QP reduction here is dominated by trapping, which can be expressed as {$s_{\mathrm{qubit} }\cdot x_{\mathrm{QP,qubit} }$}. Thus, we can obtain the evolution of QP density near the qubit
\begin{equation}
\begin{aligned}
x_{\mathrm{QP,qubit} } & = \int_{-{t_\mathrm{SFQ}} +t_{\mathrm p}}^{t_{\mathrm R}}\left ( g_{\mathrm{qubit} }- s_{\mathrm{qubit}}\cdot x_{\mathrm{QP,qubit} }  \right ) \,\mathrm{d}t\\
& {= e^{-s_\mathrm{qubit}t_\mathrm{R}} \int_{-t_\mathrm{SFQ}+t_{\mathrm p}}^{t_{\mathrm R}}\left ( e^{s_\mathrm{qubit}{t}}\cdot g_{\mathrm{qubit} }  \right ) \,\mathrm{d}t}.
\end{aligned}
\label{equ6}
\end{equation}
\noindent The SFQ-circuit-induced QP density in qubit vicinity calculated according to Eq.\;(\ref{equ6}) is shown by the blue dashed line of Fig.\;\ref{fig13}, where $\,r=1/\left ( 42\,\mathrm{ns} \right) ,\; s_{\mathrm{qubit} }=1/\left ( 9.1\,\mathrm{\mu s} \right),\;
t_{\mathrm{p} }=7.9\,\mathrm{\mu s} ,\;\alpha=1.04\times 10^{-2}$, and$\;
g_{_\mathrm{SFQ} }=2.16\times 10^2 {\,\mathrm{s^{-1}}}$. {Considering that the $x_{\mathrm{QP,qubit}}$ calculated in this model is an instantaneous value rather than an average value over a period of time in the measurement, the parameters extracted by this model are for qualitative reference.} The calculated QP density evolution trend near the qubit agrees with the density extracted from the qubit relaxation times.


\providecommand{\noopsort}[1]{}\providecommand{\singleletter}[1]{#1}%

\end{document}